\documentclass[prl,twocolumn,showpacs,preprintnumbers,floatfix,amsmath,amssymb,amsfonts]{revtex4}

\usepackage{graphicx}
\usepackage{dcolumn}
\usepackage{bm}
\usepackage{natbib}

\newcommand{\tint}{{\textstyle\int}}

\begin{document}

\title{Phase Diffusion and Lamb-Shift-Like Spectrum Shift in Classical Oscillators}

\author{Xiaofeng Li}
\author{Wenjiang Zhu}
\author{Donhee Ham}
 \email{donhee@seas.harvard.edu}
 \homepage{http://www.seas.harvard.edu/~donhee}
\affiliation{%
School of Engineering and Applied Sciences, Harvard University \\
33 Oxford Street, Cambridge, MA 02138
}%

\date{\today}

\begin{abstract}
The phase diffusion in a self-sustained oscillator, which produces oscillator's spectral linewidth, is
inherently governed by a nonlinear Langevin equation. Over past 40 years, the equation has been treated with linear approximation, rendering the nonlinearity's effects unknown. 
Here we solve the nonlinear Langevin equation using the perturbation method borrowed from quantum mechanics, and reveal the physics of the nonlinearity: slower phase diffusion (linewidth
narrowing) and a surprising oscillation frequency shift that formally corresponds to the Lamb shift in quantum electrodynamics.
\end{abstract}

\preprint{arXiv:0908.2214}

\maketitle

\section{\label{sec:intro}Problem Statement}
\vspace{-2mm}
Self-sustained oscillators are a major research subject in the broad
context of science and technology. Electronic oscillators
\cite{Van-der-Pol,Andress}, masers/lasers \cite{Townes,Maiman},
neural circuits \cite{Glass}, and human circadian rhythms
\cite{Czeisler} are examples from electronics, optics, and physiology. Such oscillators sustain oscillations on limit cycles by
compensating parasitic energy loss \cite{Van-der-Pol}.
Noise disturbs oscillation, causing amplitude and phase errors.
The amplitude error that puts oscillation \textit{off} the limit cycle is constantly
corrected by oscillator's tendency to return to its limit cycle.
By contrast, the phase error \textit{on} the limit cycle accumulates
without bound, as no mechanism to reset phase exists. Thus phase undergoes Brownian motion, diffusing along the
limit cycle. The oscillator signal with frequency $\omega_0$ may then be written as
$v(t)=v_{0}\cos[\omega_{0}t+\phi(t)]$ ignoring the amplitude error. $\phi(t)$ is the phase diffusion. Due to $\phi(t)$, $v(t)$'s spectrum
is broadened about $\omega_0$. The phase diffusion and
spectral broadening are among the most essential aspects of oscillators' dynamics and quality.

\begin{figure}[b]
\includegraphics[width=0.47\textwidth]{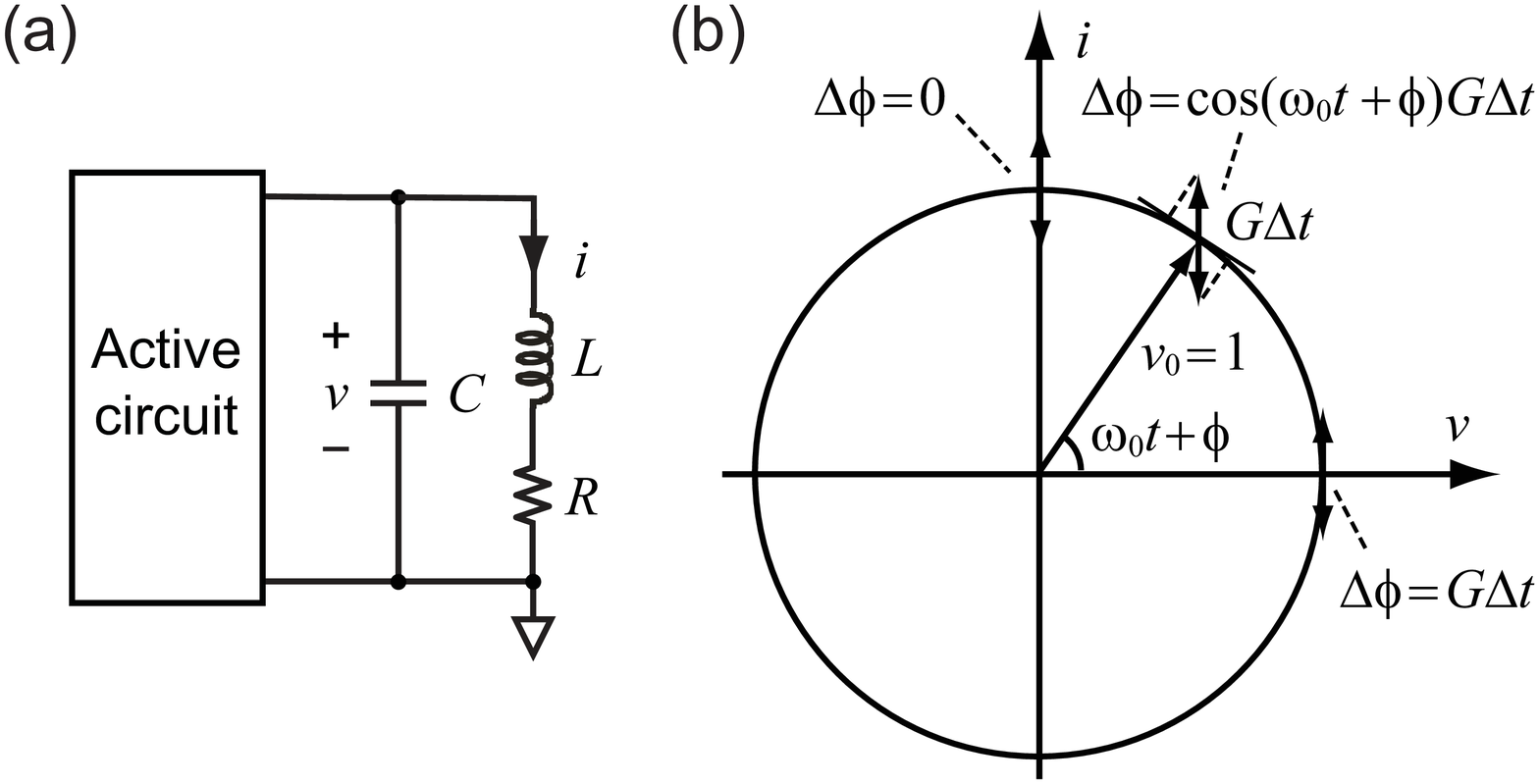}
\caption{\label{fig:phase-error} \textbf{(a)} Circuit diagram of
a self-sustained $LC$ oscillator. \textbf{(b)} Limit cycle and state-dependent phase error.}
\end{figure}

The phase diffusion due to the disturbance by a Gaussian white noise $G(t)$ is governed by an
inherently nonlinear Langevin equation \cite{Lax,Stratonovich}:
\begin{equation}
d\phi/dt=\cos(\omega_{0}t+\phi)G(t). \label{eq:original-SDE}
\end{equation}
This is the standard Langevin equation for Brownian motion, except the
periodic modulation $\cos(\omega_0 t + \phi)$ that depends on the
oscillator's state, $\omega_{0}t+\phi(t)$. This state-dependent
modulation is a hallmark property of the phase diffusion \cite{Lax,Stratonovich}: it arises, as the phase error for a given disturbance varies with where the disturbance occurs on the limit cycle. This is
shown in Fig.~\ref{fig:phase-error} for an $LC$
oscillator. During the time interval $\Delta t$, current $i$
in the $LC$ tank is disturbed by noise $G\Delta t$
produced by parasitic resistor $R$. The resulting phase error $\Delta \phi$ is a projection of $G\Delta t$ along the tangential
direction of the limit cycle, which depends on the limit cycle position. The modulation can be a general $\omega_0$-periodic function, for the
noise intensity can also vary with the oscillator's state and the limit cycle can assume a non-circular shape, but the cosine modulation captures the essence.

When $\phi$ diffuses according to (\ref{eq:original-SDE}),
probability density $p(\phi,t)$ evolves
according to the Fokker-Planck equation, $\partial p/\partial t =(\partial^2/\partial
\phi^2) [(\sigma^{2}/{2})\cos^{2}(\omega_{0}t+\phi) p]$. 
Here $\sigma^2$ is $G(t)$'s power spectral density: $\langle G(t_1)G(t_2) \rangle=\sigma^2
\delta(t_1-t_2)$, where $\langle \cdot \rangle$ is ensemble average. The instantaneous phase diffusion rate,
$\sigma^{2}\cos^{2}(\omega_{0}t+\phi)/2$, identified in the Fokker-Planck equation, depends on the oscillator
state. Its time average, after ignoring the nonlinear $\phi$-dependence, is $D \equiv \sigma^{2}/4$, which gives a sense of the average diffusion rate (the exact value can differ from $D$ due to the $\phi$-dependence, as seen later). $D$ and $\omega_0$ are oscillator's two characteristic frequencies.

Due to the nonlinear $\phi$-dependence, (\ref{eq:original-SDE}) has no closed-form solution.
In 1960s, phase diffusion was studied in
the low-noise regime, $D \ll \omega_0$, to which electronic
oscillators and lasers belong \cite{Lax,Stratonovich,Scully}.
These works dropped the nonlinear $\phi$-dependency in (\ref{eq:original-SDE}), which allowed time-averaging of the remaining modulation
to reduce (\ref{eq:original-SDE}) to
\begin{equation}
d\phi/dt \approx (1/\sqrt{2}) G(t). \label{eq:simplified-SDE}
\end{equation}
This linear approximation works, as for $D \ll \omega_0$ many
cycles of oscillation occur before phase diffuses appreciably.
The corresponding simplified Fokker-Planck equation gives a
normally distributed $p(\phi,t)$ with $\bigl \langle \phi^2(t) \bigr \rangle = 2D t$. This 
diffusion with rate $D$ leads to the well-known Lorentzian
spectrum with half-width $D$ at half-power \cite{Lax}.

For over 40 years, however, the general nonlinear Langevin equation (\ref{eq:original-SDE}) has escaped solution, and a couple of fundamental questions on phase diffusion have remained unanswered. First, while the simplified model (\ref{eq:simplified-SDE}) is valid in the low-noise regime $D \ll \omega_0$ (electronic oscillators, lasers) \cite{Lax}, the lack of solution to the general equation (\ref{eq:original-SDE}) has obscured when/how
the simplified model fails, shadowing confidence of electronics and laser community in the simplified model. Second, in the high-noise regime $D \sim \omega_0$ where the nonlinear $\phi$-dependency cannot be
ignored, what are the physical effects of the nonlinearity? Not only is this in itself a fundamental question on phase diffusion, but the answer can be useful in studying timing accuracy of low-frequency high-noise oscillators, {\it e.g.}, neural circuits \cite{Glass}, circadian rhythms \cite{Czeisler}.

Here we address this long-standing problem by solving the nonlinear Langevin dynamics (\ref{eq:original-SDE}) (Sec. 2). We thus answer both questions above (Sec. 3), drawing the boundary of the linear model and revealing the physics of the nonlinearity: slower phase diffusion (narrower spectrum) and a surprising oscillation frequency shift. Also surprising is the formal correspondence of the frequency shift in classical oscillators to the Lamb shift in quantum electrodynamics (Sec. 3); this finding highlights, with a widely-used physical system as a concrete example, the intimate link between quantum and stochastic systems known from works by Nelson \cite{Nelson} and Feynman \cite{Feynman}.

\vspace{-3mm}
\section{\label{sec:analytical}Solution via Feynman-Kac Equation}
\vspace{-2mm}

The power spectral density of $v(t)=v_{0}\cos(\omega_{0}t+\phi)$, a measurable quantity capturing phase diffusion, is Fourier transform of the autocorrelation $R_v(\tau) = \bigl\langle v(t)v(t+\tau) \bigr\rangle$. Thus our task is to compute $R_v$. Our approach is to relate $R_v$ to a function (which we call $f$) governed by Feynman-Kac equation, and solve it using the perturbation method borrowed from quantum mechanics.

We deal with the total phase $\psi(t)
\equiv \omega_{0}t+\phi(t)$. The nonlinear Langevin equation for $\psi(t)$, matched to (\ref{eq:original-SDE}), is:
\begin{equation}
d\psi/dt=\omega_{0}+\cos\psi \cdot G(t); \label{eq:psi-SDE}
\end{equation}
and the autocorrelation with $v(t)=v_{0}\cos\psi(t)$ reads
\begin{equation}
R_{v}(\tau) = \tfrac{1}{2}v_{0}^{2}\text{Re}\bigl[\bigl\langle
e^{i[\psi(t+\tau)-\psi(t)]}\bigr\rangle +\bigl\langle
e^{i[\psi(t+\tau)+\psi(t)]}\bigr\rangle \bigr].
\label{eq:auto-expansion}
\end{equation}
Here $\bigl\langle e^{i[\psi(t+\tau)\mp\psi(t)]} \bigr \rangle
= \bigl\langle e^{\mp i\psi(t)} \bigl\langle e^{i\psi(t+\tau)}|
\psi(t) \bigr\rangle \bigr\rangle_{\psi(t)}$ by conditional probability. Let $f(x,\tau) \equiv \bigl\langle e^{i\psi(t+\tau)}|\psi(t)=x \bigr\rangle$ be the ensemble average of $e^{i\psi(t+\tau)}$, given $\psi(t)=x \in[0, 2\pi]$, and $p(x)$ be the stationary probability density of $\psi(t)=x$ for $t$ large enough. Then we have
\begin{equation}
\bigl\langle e^{i[\psi(t+\tau)\mp\psi(t)]} \bigr\rangle =
\tint_{0}^{2\pi}f(x,\tau)e^{\mp ix}p(x)dx.
\label{eq:1st-term}
\end{equation}

We have reduced computation of $R_v$ to that of $f(x,\tau)$ and $p(x)$. Compared to $f$, $p$ is easier to calculate ($p$ has a closed-form expression \cite{Supplemental}), and has no major physical effects as seen later, so we focus on $f$. $f$ decays with $\tau$, for an ensemble of oscillators with the same initial phase $x$ dephase over time due to phase diffusion (decay rate = dephase/diffusion rate), thus, $f$ captures phase diffusion and is physically meaningful. Mathematically, given (\ref{eq:psi-SDE}), $f$ satisfies the following Feynman-Kac equation for $\tau>0$:
\begin{equation}
\label{eq:Feynman-Kac}
\frac{\partial f}{\partial \tau} = \underset{H_0}{\underbrace{\left[\omega_0 \frac{\partial}{\partial x} + D\frac{\partial ^2}{\partial x^2}\right.}} + \underset{H^{\prime}} {\underbrace{\left. D \cos 2x \frac{\partial ^2}{\partial x^2}\right]}} f.
\end{equation}
This does not contain $t$, which is why $t$-dependency was dropped in $f$ and $R_v$. The remaining job is to solve (\ref{eq:Feynman-Kac}). It has no closed-form solution due to the $H^{\prime}$ term, whose origin is the nonlinear $\psi$-dependency in the Langevin equation (\ref{eq:psi-SDE}). As (\ref{eq:Feynman-Kac}) has the same form as Schr\"{o}dinger equation, we solve (\ref{eq:Feynman-Kac}) using the perturbation method of quantum mechanics, separating it into unperturbed ($H_0$) and perturbed ($H^{\prime}$) terms. When $D \ll \omega_0$, $H^{\prime}$ is negligible, corresponding to the simplified model (\ref{eq:simplified-SDE}).

Let $\epsilon_n$ and $g_n(x)$ be the $n$-th eigenvalue and eigenstate of (\ref{eq:Feynman-Kac}): $(H_0+H^\prime)g_n = \epsilon_n g_n$. Quantization $n= \pm 1, \pm 2, ...$ is due to periodic boundary condition $g_n(0)$=$g_n(2\pi)$. $\epsilon_n$ and $g_n$ can be obtained via the perturbation technique. As $g_n$ evolves as $e^{\epsilon_n \tau} g_n$, $f(x, \tau) = \sum_n{c_n e^{\epsilon_n \tau} g_n(x)}$. Coefficients $c_n$ are set by initial condition $f(x, 0) = e^{ix}$. 

The unperturbed equation $H_0 g_n = \epsilon_n g_n$ with the periodic boundary condition yields the 0th-order $g_n$ and $\epsilon_n$:
\begin{equation}
\label{eq:eigen0}
g_n^{(0)}(x) = e^{inx} \text{ ; } \epsilon_n^{(0)} = in\omega_0 - n^2D.
\end{equation}
Each eigenstate is a harmonic mode of oscillation; its frequency $n\omega_0$ and decay rate $n^2D$ are Im[$\epsilon_n^{(0)}$] and Re[$\epsilon_n^{(0)}$]. In this unperturbed case, initial condition $f(x,0)=e^{ix}$ is an eigenstate, $g_1^{(0)}(x)$, of $H_0$, so $f(x,\tau)=e^{\epsilon_1^{(0)} \tau} g_1^{(0)}(x) = e^{-D\tau} e^{i(\omega_0 \tau + x)}$ with no other eigenstates excited.

For higher-order solutions, we obtain matrix elements of $H^{\prime}$ with the unperturbed eigenstates $g_n^{(0)}(x)$ as basis:
\begin{equation}
H_{mn}^{\prime} \equiv \tfrac{1}{2\pi} \tint_0^{2\pi}{g_m^{(0)\ast}H^{\prime}g_n^{(0)}dx} = -(n^2D/2)\delta_{m,n\pm 2}. \label{eq:matrix-element}
\end{equation}
This off-diagonal matrix alters eigenstates $g_n$ and eigenvalues $\epsilon_n$ of $H_0+H^\prime$ from the unperturbed ones, whose calculation to the lowest order we describe now. In the lowest-order correction, $g_n^{(0)}$ is coupled only with $g_{n \pm 2}^{(0)}$ given (\ref{eq:matrix-element}), thus, $g_n=g_n^{(0)}+a_{n,n+2}g_{n + 2}^{(0)}+a_{n,n-2}g_{n - 2}^{(0)}$ with $a_{n,n\pm 2} = H^\prime_{n\pm2,n}/(\epsilon_n^{(0)}-\epsilon_{n\pm2}^{(0)})$ obtained from the perturbation formula of quantum mechanics. Also $\epsilon_n$, or the frequency and decay rate, is altered due to $g_n^{(0)}$'s coupling with $g_{n \pm 2}^{(0)}$ in its lowest-order correction:
\begin{equation}
\epsilon_n= \epsilon_n^{(0)}+\sum_{m = n \pm 2}\frac{H^{\prime}_{mn}H^{\prime}_{nm}} {\epsilon_n^{(0)}-\epsilon_m^{(0)}},
\label{eq:correction}
\end{equation}
With $g_n$ and $\epsilon_n$ available to the lowest order, we can find $f(x,\tau)$ to the same order by using the initial condition, $f(x,0) = e^{ix}$. As $e^{ix} = g_1^{(0)}(x)$ is not an eigenstate of $H_0+H^\prime$, it can be expanded in terms of $g_n$'s, the eigenstates of $H_0+H^\prime$; to the lowest order, we find $e^{ix}=c_1 g_1 + c_{-1} g_{-1} + c_3 g_3$  with $c_1=1$, $c_{-1}=D/(i4\omega_0)$, $c_3=D/(-i4\omega_0+16D)$ ($|c_1|$ is far larger than $|c_{-1}|$ and $|c_{3}|$ for $D \ll \omega_0$, as $f(x,0)=g_1^{(0)}(x)$ is still close to $g_1(x)$ for  $D \ll \omega_0$) \cite{Supplemental}. We can then readily write $f(x,\tau)=c_1 e^{\epsilon_1 \tau} g_1(x) + c_{-1} e^{\epsilon_{-1} \tau} g_{-1}(x)+c_3 e^{\epsilon_3 \tau} g_3(x)$ to the lowest order with no other eigenstates excited. The $e^{\epsilon_{\pm 1} \tau}$ terms dominate as they have slowest decay rate ($\sim n^2 D = D$). Ignoring the $e^{\epsilon_3 \tau}$ term with far higher decay rate ($\sim n^2D = 9D$) and using $\epsilon_n=\epsilon_{-n}^\ast$, $g_n=g_{-n}^\ast$,
\begin{equation}
f(x,\tau)\approx c_1 e^{\epsilon_1 \tau} g_1(x) + c_{-1} e^{\epsilon_{1}^\ast \tau} g_{1}^\ast(x).
\label{eq:f-2terms}
\end{equation}

We can extend the calculation to any higher order \cite{Supplemental}. In higher orders, $f$ contains more excited eigenstates, but (\ref{eq:f-2terms}) still holds (with more accurate $\epsilon_{1}$, $g_{1}$, and $c_{\pm1}$), given the slowest decay of the two terms (secondarily, $|c_1|$ is larger than other $|c_n|$'s due to the initial condition).

Using (\ref{eq:f-2terms}) in (\ref{eq:1st-term}) yields $R_v$. To the lowest order, it is:
\begin{equation}
R_v(\tau) \approx \tfrac{v_{0}^{2}}{2}e^{-(D-\alpha) |\tau |}
\left[\cos(\omega_{0}+\beta)\tau-\tfrac{D}{2\omega_{0}}
\sin(\omega_{0}+\beta)|\tau|\right]. \label{eq:auto-expr}
\end{equation}
Here $\alpha \equiv \text{Re}[\epsilon_1-\epsilon_1^{(0)}]$ and $\beta \equiv \text{Im}[\epsilon_1-\epsilon_1^{(0)}]$ become the lowest order correction of $\epsilon_1$ according to (\ref{eq:correction}):
\begin{subequations}
\label{eq:alpha-beta}
\begin{align}
\alpha & = (9/2) [D^2/(16D^2+\omega_{0}^{2})]D; \label{eq:alpha} \\
\beta & = (D/8) [ 9D\omega_{0}/ (16D^2+\omega_{0}^{2}) - D/\omega_0 ], \label{eq:beta}
\end{align}
\end{subequations}
which represent decay rate reduction and frequency shift, caused by $g_1^{(0)}$'s coupling with $g_{-1}^{(0)}$, $g_3^{(0)}$. For (\ref{eq:auto-expr}), we approximate $p(x)$ of (\ref{eq:1st-term}) as uniform, which is accurate for $D\ll \omega_0$. The exact $p(x)$, found in closed form by solving the {\it stationary} Fokker-Planck equation corresponding to (\ref{eq:psi-SDE}), slightly alters coefficients of the terms in $R_v$, while the rest, including the physically important $\alpha$ and $\beta$ that are from $f(x,\tau)$, remain the same \cite{Supplemental}. In higher orders, the functional form of (\ref{eq:auto-expr}) is maintained, as (\ref{eq:f-2terms}) is still used, while the coefficients, $\alpha$, and $\beta$ are more accurate due to higher-order corrections of $\epsilon_1$ and $g_1$.

\vspace{-3mm}
\section{\label{physics} Slow Diffusion and Frequency~Shift}
\vspace{-2mm}

The autocorrelation $R_v$ in (\ref{eq:auto-expr}) captures physical consequences of the nonlinear Langevin dynamics (\ref{eq:original-SDE}). First, the decay rate $D-\alpha$, or the linewidth of the power spectrum corresponding to $R_v$, is the phase diffusion (dephase) rate, which is smaller than the rate $D$ obtained from the approximate linear model (\ref{eq:simplified-SDE}). Second, the oscillation frequency $\omega_0 + \beta$ seen from (\ref{eq:auto-expr}) is surprisingly shifted from the natural oscillation frequency $\omega_0$ by $\beta$. In the low-noise regime ($D \ll \omega_0$) where the lowest-order $\alpha$ and $\beta$ in (\ref{eq:alpha-beta}) are highly accurate, $\alpha /D \propto (D/\omega_{0})^{2} \ll 1$ and $\beta/\omega_0\propto (D/\omega_{0})^{2} \ll 1$, thus the true nonlinear dynamics (\ref{eq:original-SDE}) and simplified model (\ref{eq:simplified-SDE}) quadratically converge as $D$ decreases. In contrast, if $D$ increases towards $\omega_0$, $\alpha$ and $\beta$ grows appreciable: for $D=\omega_0$, $\alpha/D=0.26$ from (\ref{eq:alpha}), and $\alpha/D=0.31$ by the $n$-th order calculation until convergence; the $n$th-order $\beta$ is 8\% of $\omega_0$. The analysis decisively justifies the linear model (\ref{eq:simplified-SDE}) for $D\ll \omega_0$, finding its boundary set by the quadratic convergence, thus, enhancing confidence in the linear model used in lasers and electronic oscillators. It has been also found that the physical effects of the nonlinearity are the slower diffusion and frequency shift, and they are conspicuous in the high-noise regime ($D \sim \omega_0$).

The slower phase diffusion may be understood as follows. The
instantaneous diffusion rate $2D\cos^{2}(\omega_{0}t+\phi)$ is zero
when $\phi =-\omega_{0}t+n\pi/2$ with odd integer $n$. These
$\phi$-points, which change with time, act as barriers that prevent
phase diffusion across them. When $\omega_{0}$ is small, the $\phi$
barrier points tend to move slowly or be more fixated, making their
effects more potent. In the extreme case of $\omega_{0} =0$, $\phi$
cannot diffuse to the region $(\pi/2,3\pi/2)$ starting from
$\phi(t=0)\in [-\pi/2,\pi/2]$. Overall, this barrier action leads to
slower phase diffusion.

\begin{figure}[b]
\includegraphics[width=0.33\textwidth]{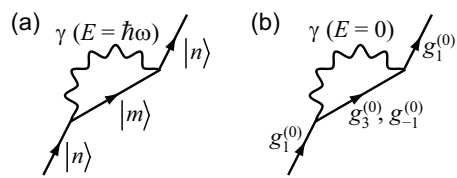}
\caption{\label{fig:feynman-diagram} Feynman diagrams of the lowest-order process for
\textbf{(a)} the Lamb shift and \textbf{(b)} our frequency shift.}
\vspace{-5mm}
\end{figure}

The frequency shift, no matter how small, is surprising and interesting from the perspective of time-domain dynamics, as noise disturbs phase in positive and negative directions with equal probability and no such shift is expected. In our calculation using eigenstates, the frequency shift comes out as part of the eigenvalue shift. It arises from the interactions among the harmonic modes ($g_n^{(0)}$) caused by the $D\cos2x$ term ($H^\prime$) in (\ref{eq:Feynman-Kac}), or the matrix element $H^\prime_{mn}$, as seen in (\ref{eq:correction}). The calculation reveals that both the noise fluctuations captured by $D$ and nonlinear modulation captured by $\cos 2x$ are the origin of $H^\prime_{mn}$, hence, indispensable for the frequency shift.

Also interesting is the formal correspondence of this frequency shift in classical oscillators to the Lamb shift in quantum electrodynamics. In our frequency shift, noise fluctuations ($D$) and the nonlinear modulation ($\cos 2x$) play the role of vacuum fluctuations ($\textbf{A}$) and the non-diagonal dipole operator ($\textbf{p}$) in the Lamb shift. The latter two enter the interaction Hamiltonian $H^{\prime} = -(e/2mc) (\textbf{p} \cdot \textbf{A} + \textbf{A} \cdot \textbf{p})$, which couples different atomic states to produce the Lamb shift \cite{Senitzky}:
\begin{equation}
\Delta E_n = \sum_{m\neq n}\sum_{\text{photon}}\frac{|H^{\prime}_{mn}|^2}{E_n - E_m - \hbar\omega}. \label{eq:lamb-shift}
\end{equation}
The summation is over both atomic states and virtual photon (vacuum fluctuation) states. Notice its similarity to (\ref{eq:correction}), except the missing $\hbar\omega$ term in (\ref{eq:correction}). This is sensible, as the ``virtual photons" in our case has zero energy, for phase disturbance does not incur energy exchange. Figure \ref{fig:feynman-diagram} illustrates the correspondence using Feynman diagrams of the lowest-order process.

Atomic level shifts {\it per se} are common in quantum mechanics, but what makes Lamb shift unique is its origination from fluctuations (our focus in Lamb shift is its physical origin rather than its computation using renormalization and relativistic corrections). This is why our frequency shift, which also occurs due to fluctuations, uniquely corresponds to Lamb shift. This finding can be put in the fundamental context of the intimate formal link between stochastic and quantum dynamics, established by Nelson \cite{Nelson}, Feynman \cite{Feynman}: the link arises as both deal with the evolution of probability-related functions. In this context, it is interesting to seek for a real physical system as a stochastic analogue of Lamb shift. The phase diffusion with the unique natural nonlinearity in classical oscillators offers one such stochastic analogue.

\begin{figure}[t]
\includegraphics[width=0.47\textwidth]{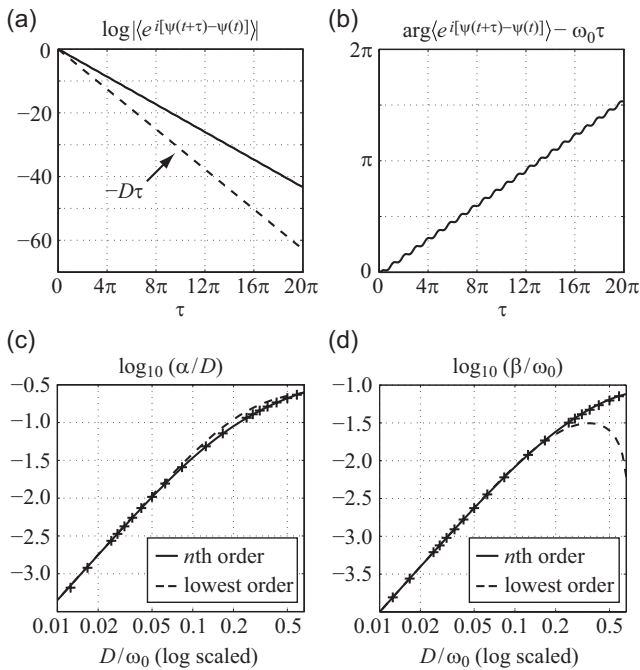}
\vspace{-2mm}
\caption{\label{fig:alpha-beta} \textbf{(a),(b)} Numerical $\bigl\langle e^{i[\psi(t+\tau) - \psi(t)]} \bigr\rangle$ for $D = \omega_0$; \textbf{(c),(d)} Numerical $\alpha$, $\beta$ (cross marks) vs. analytical $\alpha$, $\beta$ to the lowest (dashed lines) and the $n$-th order (solid lines).}
\vspace{-3mm}
\end{figure}

\begin{figure}[t]
\includegraphics[width=0.47\textwidth]{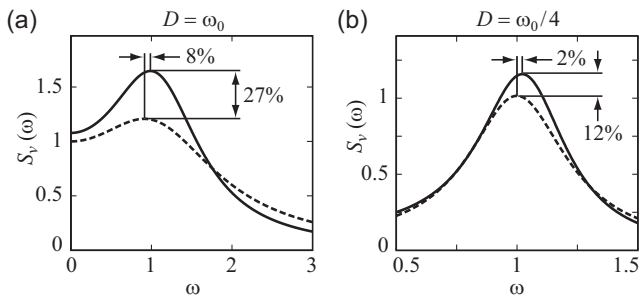}
\vspace{-2mm}
\caption{\label{fig:spectrum} Power spectra from Monte-Carlo simulations of (\ref{eq:original-SDE}) (solid) and (\ref{eq:simplified-SDE}) (dashed). $\omega_0=1$: \textbf{(a)} $D=\omega_0$; \textbf{(b)} $D = \omega_{0}/4$.}
\vspace{-5mm}
\end{figure}

\section{\label{sec:numerical}Numerical Calculations}
\vspace{-3mm}
To verify our solution to the Feynman-Kac equation, we solve
(\ref{eq:Feynman-Kac}) using the finite difference method, and
calculate $\bigl\langle e^{i[\psi(t+\tau)-\psi(t)]} \bigr \rangle$ of (\ref{eq:1st-term}).
The nearly linear relations of $\log |\bigl\langle e^{i[\psi(t+\tau) - \psi(t)]} \bigr\rangle|$ and $\arg \bigl\langle e^{i[\psi(t+\tau) - \psi(t)]} \bigr\rangle$ to $\tau$ for $D = \omega_0$ in
Figs.~\ref{fig:alpha-beta}(a),(b) confirm that a single exponential
$e^{-(D-\alpha)\tau}e^{i(\omega_{0}+\beta)\tau}$ indeed dominates
the autocorrelation. The location of $\log|\bigl\langle e^{i[\psi(t+\tau) - \psi(t)]} \bigr\rangle|$
above the line of $-D\tau$, and the positive slope of
$\arg \bigl\langle e^{i[\psi(t+\tau) - \psi(t)]} \bigr\rangle - \omega_{0}\tau$, confirm the reduced
diffusion rate and frequency shift.
Figures~\ref{fig:alpha-beta}(c),(d) plot $\alpha$ and $\beta$
from our analytical calculations to the lowest order
(\ref{eq:alpha-beta}) and the $n$th order (up to convergence at
$n \sim 10$ \cite{Supplemental}), against their numerically obtained values. Our analysis agrees well with the numerical results: the
discrepancy between (\ref{eq:alpha-beta}) and the numerical result at large $D$ is readily corrected by higher-order calculations. Note
also that $\alpha/D$ and $\beta / \omega_{0}$ diminish as
$(D/\omega_{0})^{2}$ with decreasing $D$ as
shown in Figs.~\ref{fig:alpha-beta}(c),(d).

For further confirmation, we Monte-Carlo simulate the
nonlinear Langevin dynamics (\ref{eq:original-SDE}). The
resulting power spectra $S_v(\omega)$ (Fig.~\ref{fig:spectrum}) are sharper/narrower than the simplified model, confirming the slower
phase diffusion: the rate reduction is 31\% in our analysis and 27\% in
the Monte-Carlo simulation for $D = \omega_0$; 11\% and 12\% for
$D = \omega_0 / 4$. The spectrum peak exhibits a definitive blue shift in centre
frequency (Fig.~\ref{fig:spectrum}). The shift for $D = \omega_0 / 4$ is 2\%
in Monte-Carlo simulation, and 3\% in our analysis. For $D = \omega_0$, the spectrum peak does not exactly correspond to
$\omega_{0}+\beta$ due to the large $D$, thus, reading the spectrum requires caution, but the extracted
$\beta$, which is a good representation of the frequency
shift, matches well between the simulation (8\%) and
our analysis (8\%).


%
\end{document}